\documentclass[preprint,showpacs,amsmath,amssymb]{revtex4}

\usepackage{graphicx}
\usepackage{dcolumn}
\usepackage{bm}
\usepackage{amsmath}

\begin{document}

\title{The Topology of Music Recommendation Networks}

\author{Pedro Cano}
\email{pcano@iua.upf.es}
\author{Oscar Celma}
\author{Markus Koppenberger}
\affiliation{
Music Technology Group, Universitat Pompeu Fabra,
Ocata 1, 08003 Barcelona, Spain
}
\author{Javier M. Buld\'u}
\email{javier.martin-buldu@upc.edu}
\affiliation{
Departament de F\'{\i}sica i Enginyeria Nuclear, Universitat
Polit\`ecnica de Catalunya, Colom 11, E-08222 Terrassa, Spain
}

\date{\today}

\begin{abstract}
We study the topology of several music recommendation networks, which rise from relationships
between artist, co-occurrence of songs in playlists or experts' recommendation. 
The analysis uncovers the emergence of complex network phenomena in this kind of
 recommendation networks, built considering artists as nodes and their resemblance as links.
We observe structural properties that provide some hints on navigation and possible 
optimizations on the design of music recommendation systems. Finally, the analysis derived from existing music knowledge sources provides a deeper understanding of the human music
 similarity perceptions. 
\end{abstract}


\maketitle

\noindent
{\bf

Music is ubiquitous in human societies. Music generates communities of
musicians \cite{lim03,gleis}, communities of listeners. Nevertheless, the way music links people is certainly diverse and sometimes unexpected. In this work we focus on networks where musicians (or bands) are the fundamental nodes 
and are linked to others if they perform or compose similar music. 
This information is extracted from main on-line music recommendation systems: AllMusicGuide, MSN-Entertainment, Amazon and Launch-Yahoo!. Music recommendation systems are constructed to
 assist users to navigate through music collections, where navigation consists of guided links
among artists. When the user selects an artist, a certain number of alternative artists are suggested, which in principle should be of his/her interest.

In our study of the structure of different music recommendation systems we find characteristics which influence systems' usability. Our results show that despite some common features, such as 
small-worldness, different network characteristics exist, such as the
link degree distribution.  
We show that there exist a relation between the link degree
distribution and the construction of the networks. Networks
constructed by collaborative effort are scale-free whereas networks
with human experts supervising the links are exponential. 
This raises a discussion on the main forces driving the
 creation of the networks and hence their quality and potential uses. If preferential attachment takes
  place, as in the scale-free networks under study, the
   recommendations are biased towards popular items. On
    the other hand exponential networks are more
     faithful to the underlying music similarity.
}

\section{Introduction}

Nowadays access to music is possible by querying artists or song names ---editorial data--- or browsing recommendations generated by collaborative filtering ---i.e: recommendation systems that exploit information such as \textit{``users that bought this album also bought this album''}. 
An obvious drawback is that consumers need to know the name of the song or the artist, or an important number of consumers must have heard and rated the music. 
This situation makes it difficult for users to access and navigate through the vast amount of music composed and performed by unknown new artists, which is available on-line in an increasing number of sites. 

In this work, complex network measurements ~\cite{bar02,new03} are used to analyse the topology of networks underlying 
the main music recommendation systems. 
The properties that emerge raise a discussion on the underlying forces driving collaborative systems and expert-guided networks. 
We can also obtain some hints about how much of the network structure is due to content similarity and how much to the self-organization of the network. 
Therefore, it can shed new light on the design and validation of music similarity measures
and its evaluation~~\cite{ell02}. 
Furthermore, it uncovers possible optimisations when designing music information systems, such 
as the optimal number of links between artists or the shortest path from
artist to artist. In this sense, recommendation networks can be optimized by adding (or removing) links to facilitate navigating from artist to artist in a short number of \textit{clicks}. 
Finally, we can obtain information about which artist has more links or which genres are
more extended. This kind of information may help to understand the dynamics of certain aspects
 of music evolution, e.g: how did an artist get popular or how the music genres emerged.


\section{Graph Dataset}

We have gathered information from four different music recommendation networks; AllMusicGuide ~\cite{all}, Amazon ~\cite{ama}, Launch-Yahoo! ~\cite{lau} and MSN-Entertainment ~\cite{mus}, and we have created a graph for each source, taking the {\textit``similarity''} between artists 
as the linking parameter.
A graph is constructed as follows: each node represents a music artist whereas an edge denotes a similarity among them. The decision to create a link  between two 
artists depends on the recommendation systems criterion, which can be different from network to network.
Therefore, it is important to define how links between artists are created. The main
characteristics of each network are summarized as follows:

\begin{itemize}

\item {\em MSN-Entertainment} (MSN) is a portal to access multimedia content. Music can be accessed using editorial metadata, i.e., artist name or song title, as well as navigating through music styles. Another browsing feature, the {\em SoundsLike Artists} 
allows users to navigate from artist to artist that sounds similar. 
Ratings of similarity between artists are constructed
from user contributions. It seems to follow a collaborative filtering approach ~\cite{sar01} 
to create links between artists.

\item {\em Amazon} is an on-line retailer of music, it uses item-to-item collaborative filtering to recommend albums and artists, based on consumer ratings and habits~\cite{lin03,sar01}. We study the network constructed by using similar artists' links. It is worth
noting that in the Amazon network links are indirectly created 
by users whose knowledge of the network
nodes is limited. 

\item {\em AllMusicGuide} (AMG) is a database of music content covering facts about albums an
artists where several descriptions are considered, such as styles, moods, country of origin, 
even the birth date of the artist. An editorial group, made of a substantial number
of music experts, is responsible of the addition of new nodes (artists) and their connections.
Contributions from the users of the system are also accepted, but always under
the supervision of the editorial group. In this sense, it is a network where a filtering
process has been done. AMG defines different networks, e.g.: influences, roots, performed songs by, and so on.
In this work, in order to compare with other music recommendation networks, we focus in the network of similarity between artists.
 
\item {\em Launch-Yahoo!} (Yahoo) is a music entertainment portal which among other features allows to navigate by similar artist.
No information is given about how links between artists are created. Nevertheless, as we will
see, some conclusions can be extracted from the analysis of
the network properties.  

\end{itemize}

As a general feature, all networks are directed, which means that
 an artist A (e.g. ``Oasis'')
can be similar to an artist B (e.g. ``The Beatles''), but not necessary in the opposite direction.
The number of artists ($n$) and links ($m$) of each network is summarized in Table~\ref{tab:tab01}. 



\section{Network Properties}\label{sec:networkProperties}

Before going to the data analysis, let us introduce some definitions 
and concepts that will be used in this paper. A {\em network} 
or graph is a set of nodes (also called vertices) connected 
via links (also called edges). Networks connected by directed 
links are called {\em directed networks} while networks connected 
by undirected edges are called {\em undirected networks}. 
In order to take a decision 
about the network structure we have measured the 
following graph parameters:

\begin{itemize}

\item {\em Degree}: The degree $k_i$ of a vertex $i$ is the number of connections of that vertex and $\left< k
\right>$ is the average of $k_i$ over all the vertices of the network. 

\item {\em Degree distribution}: The degree distribution $P(k)$ is the proportion of nodes that have a degree $k$. The shape of the degree distribution can help to identify 
they type of network:
regular networks have a constant distribution, since all nodes have the same amount
of degrees, ``random networks''~\cite{rap57,erd59} ---as described by the Erd\"os-R{\'e}nyi model---have a Poisson degree distribution and ``scale-free networks'' have power-law distributions ~\cite{bar99}. 
It is a standard practice to compute the cumulative degree distribution $P_c(k)=\sum_{k'>k}P(k')$
since it filters fluctuations of $P(k)$ which is frequently rather noisy.  
In a directed graph (all graphs under study) we
can calculate $ P^{in}(k)$ and $P^{out}(k)$ as the in and out degree 
(for incoming/outgoing links) respectively. 

\item {\em Average shortest path}: Two vertices $i$ and $j$ are connected if one can go from $i$ to $j$ following
the edges in the graph. The path from $i$ to $j$ may not be unique. The minimum path distance or {\em geodesic path} $d_{ij}$ is the shortest path distance from $i$ to $j$. The average shortest path over every pair of vertices is
\begin{equation}
\left<d\right>=\frac{1}{\frac{1}{2}n(n+1))}\sum_{i \ge j}d_{ij}
\label{distance}
\end{equation}
The maximum geodesic path between any
two vertices in the graph is known as {\em diameter}.

\item {\em Clustering coefficient}: The clustering coefficient estimates the probability that two neighboring vertices of a given vertex are neighbors themselves. In music networks, it relates to the probability that if artist A is similar to artist B and artist C, B and C are similar as well. Following~\cite{wat98} the clustering coefficient of vertex $i$ is the ratio between the total number $y_i$ of the edges connecting its nearest neighbors and the total number of all possible edges between all these nearest neighbors.
$c_i$ can be calculated following the expression (see \cite{wat98} for details):
 \begin{equation}
c_i=\frac{2y_i}{k_i\left(k_i-1\right)}
\label{clusteri}
\end{equation}
Finally, the clustering coefficient $C$ for the whole network is the average over the number of nodes $n$:
 \begin{equation}
C=\frac{1}{n}\sum_{i}c_i
\label{cluster}
\end{equation}

\end{itemize}

Table~\ref{tab:tab01} summarises the network parameters of the different graphs, which
will be analyzed in the following section.

\begin{table*}[!h]
\begin{center}
\begin{tabular}{|r|c|c|c|c|c|c|c|c|c|c|}\hline
& type       &  $n  $  &$m  $    &  $\left<k\right>$ &  $C$ &   $C_r$   & $d$ &     $d_r$&$\gamma_{in}$&$\gamma_{out}$\\
\hline \hline
    MSN
&directed     & 51,616&279,240& 5.5 & 0.54 & $1.0 \centerdot 10^{-4} $ &
 7.7  &   6.4  &          2.4$\pm$0.01& - \\
\hline
    Amazon
&directed  & 23,566  & 158,866 & 13.4 & 0.14 &      $5.7   \centerdot 10^{-4}$&
4.2 & 3.9  &2.3$\pm$0.02&2.4$\pm$0.04\\
\hline
AMG
&directed  & 29,206 &  146,882 & 8.15 & 0.20 & $ 2.8  \centerdot 10^{-4}$    & 6.2 & 4.9  & - & - \\
\hline
Yahoo
&directed  & 16,302 & 511,539 & 62.8 &0.38&    $3.8 \centerdot 10^{-3}$&
2.7&2.3 & - & - \\

\hline
\end{tabular}
\small
\caption{Summary of the network parameters, where $n$ is the number or artists, 
$m$ is the number of links, $\left<k\right>$ 
   is the average degree, 
$C$ is the clustering coefficient, $C_r$ is the clustering of the equivalent
random network, $d$ 
is the average shortest path, and  $d_r$ is the corresponding 
shortest path for the random network. The last two columns correspond to
 the exponents of the 
power-law  decay of the degree distribution for 
the incoming and outgoing links, $\gamma_{in}$,$\gamma_{out}$,  of the graphs (we show exponents
only when we found power-law decay).}
\label{tab:tab01}
\end{center}
\end{table*}
\normalsize


\section{NETWORK STRUCTURE}

A common feature appears in all networks under study, they have small-world properties ~\cite{wat98}. The average shortest path $d$ (see Table~\ref{tab:tab01}) of all graphs 
is below eight and always in the same order of magnitude as the shortest
path of a random graph with the same number of nodes and links. This indicates that
despite the high number of nodes (artists) and the sparsity of the network, 
a user can always jump from a node to any
other by a short number of jumps (i.e. links). At the same time, the clustering coefficient $C$,
is several orders of magnitude higher than that of the corresponding random network. Both
ingredients, the low shortest path and the high clustering, are the typical 
characteristics of small-world networks ~\cite{wat98}. Small-worldness is an interesting property 
for recommendations systems, since it has been suggested that humans find it easy to navigate
   in small-world networks using only local information~\cite{kle00,wat02,mou03}. 

Concerning the average degree $\left<k\right>$, we observe that it has a low value, in three of them (MSN, Amazon and AMG) but it is higher in Yahoo ($\left<k\right>$=62.7). As we will discuss in the following section, the output degree tends to be 
bounded due to the usability constraints, i.e. the number of output links has to fit on a web page length.

The cumulative degree distribution $P_{c}(k)$ (distribution of nodes with a degree
equal or higher than $k$) and specifically the way that $P_{c}(k)$ decreases allows to classify the 
small-world networks ~\cite{ama00}. With this aim, we have analyzed the $P_{c}(k)$ distribution 
for all networks and we have found differences that 
are related with the internal structure and probably the
construction mechanisms of each network.

\begin{figure}
\centering
\includegraphics[width=120mm,clip]{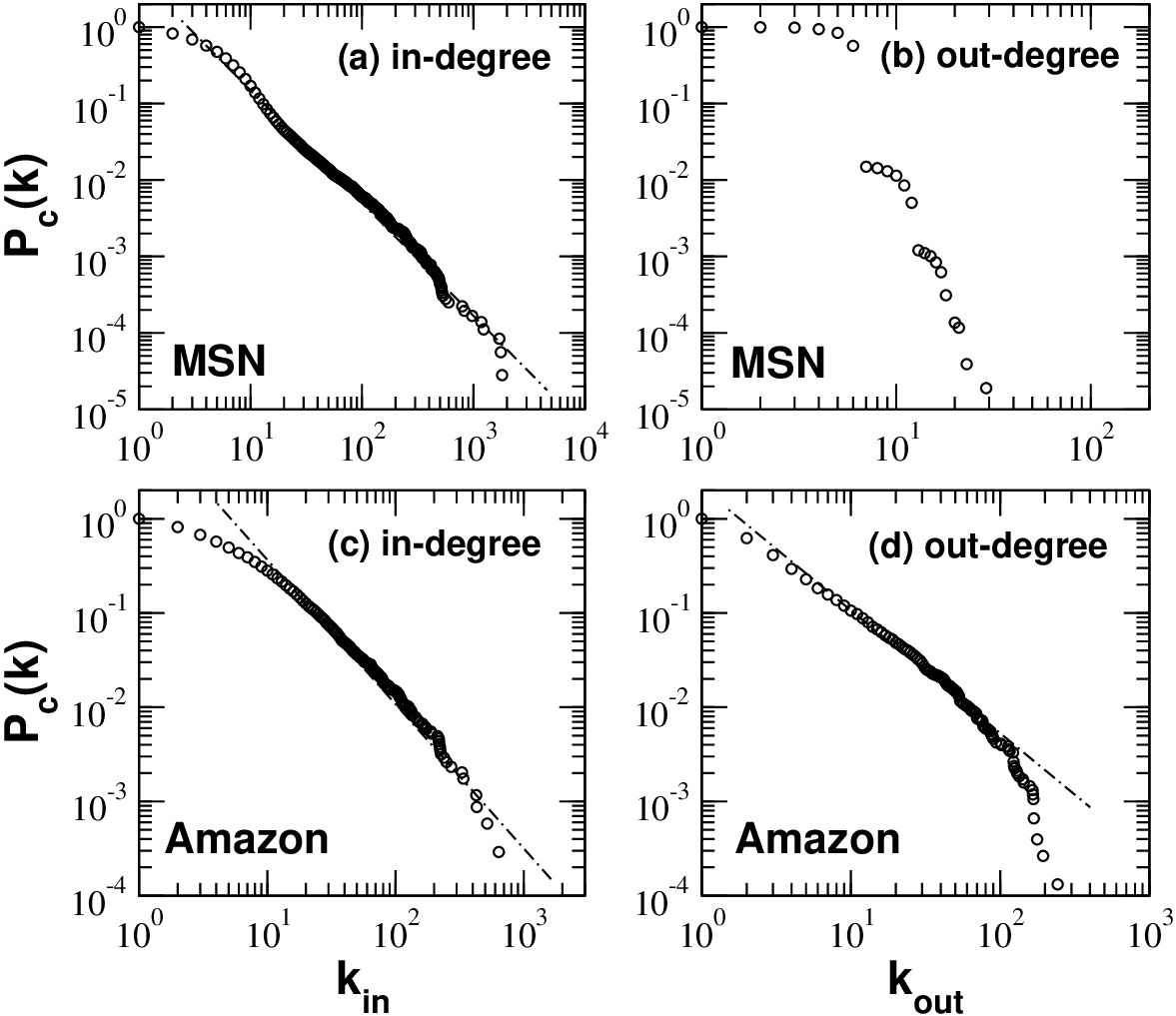}
\caption{
$P_{c}(k)$ of MSN and Amazon recommendation networks (note the log-log scale in all plots).
In the left column, the cumulative distributions $P_{c}^{in}(k)$ of 
the incoming degree $k_{in}$. The cumulative distributions $P_{c}^{out}(k)$ of the outgoing
degree $k_{out}$ are plotted in the right column.
}
\label{fig:fig01}
\end{figure}

Fig.~\ref{fig:fig01} shows the cumulative degree distribution
of the incoming (a) and outgoing (b) links of the MSN recommendation network.
It is worth noting the difference between both distributions. The incoming
degree distribution $P_{c}^{in}(k)$ refers to artists who are similar to a 
selected artist. $P_{c}^{in}(k)$ is related with 
the number of  links pointing to an artist and in a certain sense, it is an 
indicator of the influence of that artist over the others or how prototypical it is for a certain type of music. On other hand, $P_{c}^{out}(k)$ refers to the number of
output links from a given artists. The number of
outgoing connections displayed to the user is limited for practical 
reasons since networks are on-line systems where 
recommendations are shown on a web page, e.g., it would be useless and impractical to propose 2000 connections. This fact truncates the tail of the $P_{c}^{out}(k)$ distribution since not
all the similar artists are linked and reduces the expected value of $k_{max}$ \cite{kmax}
for scale-free networks.

$P_{c}^{in}(k)$ of MSN [see Fig.\ref{fig:fig01}(a)] has a power-law decay 
($P(k)\sim k^{-\gamma}$), as indicated 
by the straight line in the log-log scale. Networks with a power law decay are
called ``scale-free'' ~\cite{bar99} since we can not identify a single characteristic scale.
This kind of structure is common in small-world networks but not universal, 
and has been reported in different complex networks, such as the 
WWW ~\cite{bro00}, the network of protein interactions
~\cite{jeo01} or the telephone call graph ~\cite{abe98}.
The $P_{c}^{out}(k)$ distribution of MSN
shows that the outgoing links are limited to twelve.
Furthermore the fact that $P_c^{out}(k)$ drastically falls at $k=7$ reveals that each
artist has typically six outgoing links. This limitation has very likely been  
introduced by the system designers. It levels the outgoing links of all nodes
of the network and rules out any possibility of showing a power-law decay.

Fig.~\ref{fig:fig01} also shows the $P_{c}(k)$ of Amazon recommendation network.
$P_{c}^{in}(k)$ [Fig.\ref{fig:fig01}(c)] is quite similar to MSN, which
indicates that both networks have similar structure, at least for the incoming links. We
find again a power-law decay, which indicates that Amazon is scale-free. 
On the contrary, we find differences at $P_{c}^{out}(k)$, which keeps the power-law decay.
This is not common on this kind
of networks since it means that the outgoing links are not as limited as MSN, in fact
there are nodes with more than 100 outgoing links [see Fig.\ref{fig:fig01}(d)]. The 
absence of strong restrictions in the outgoing connections of Amazon network allows both cumulative 
distributions (incoming/outgoing degree) to have similar shapes.

In Table~\ref{tab:tab01} we have indicated the values of the power-law exponents $\gamma$ of
Fig.~\ref{fig:fig01} (see ~\cite{note}). In all cases, $\gamma$ is within the 
common range of values of previously studied scale-free networks ~\cite{new03}.   

\begin{figure}
\centering
\includegraphics[width=120mm,clip]{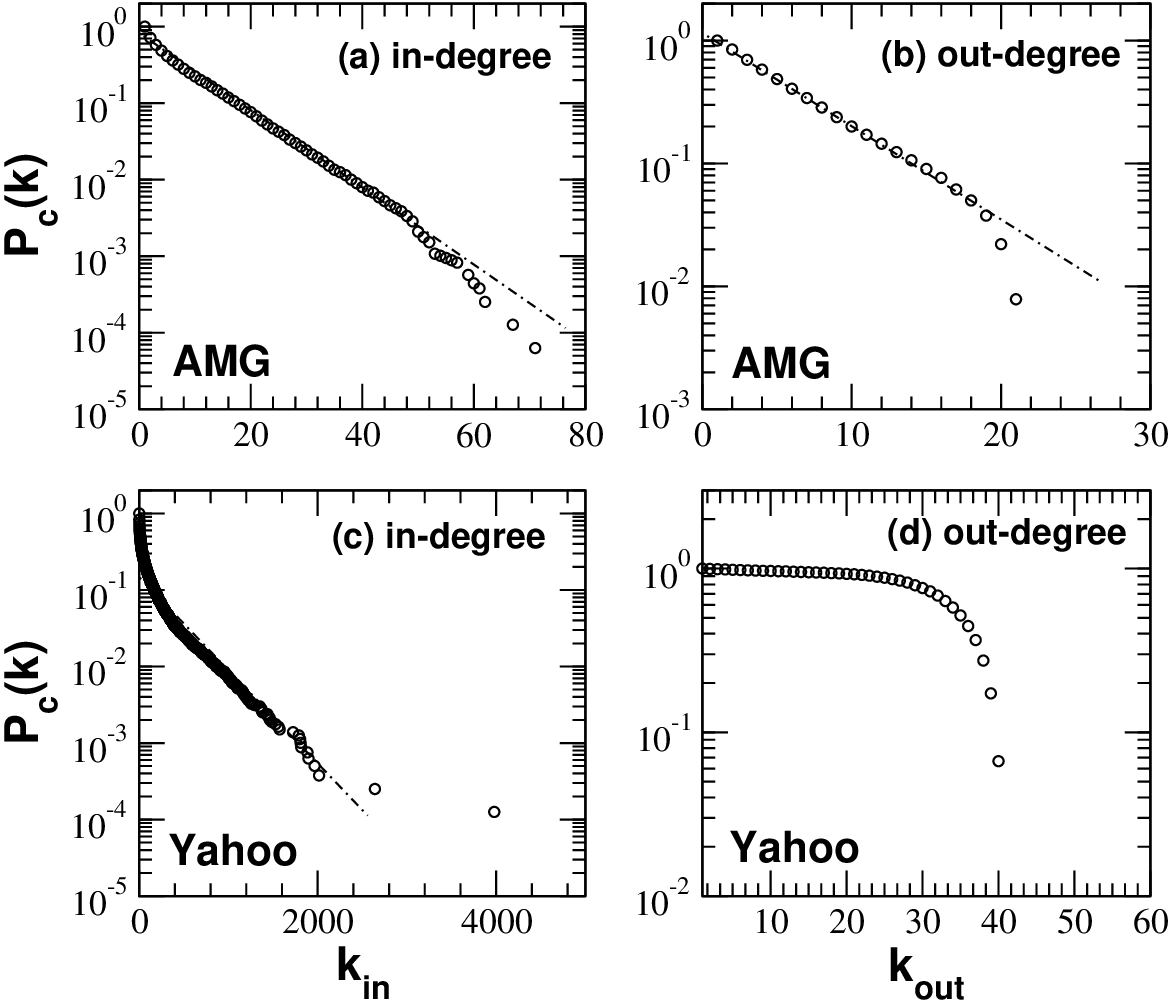}
\caption{
$P_c(k)$ of AMG and Yahoo recommendation networks, in a linear-log scale.
In the left column, the cumulative degree distributions $P_{c}^{in}(k)$ of 
the incoming links. $P_{c}^{out}(k)$ of the outgoing
links are plotted in the right column.
}
\label{fig:fig02}
\end{figure}

In Fig.~\ref{fig:fig02} we have plot the cumulative degree distributions of 
AMG and Yahoo networks, since both behave different from the previous ones.
$P_{c}^{in}(k)$ distribution of AMG [Fig.~\ref{fig:fig02}(a)] has an exponential decay  
($P(k)\sim e^{-\frac{k}{\kappa}}$)
since it follows a straight line in the linear-log scale
(note the linear scale in the horizontal axes).
$P_{c}^{out}(k)$ keeps the exponential decay [see Fig.\ref{fig:fig02}(b)].
It is interesting to note that although the outgoing
links are limited to 21, $P_{c}^{out}(k)$ still has
exponential decay, although the slope is different from that of the incoming links.

Finally, we observe how Yahoo shows some similarities with AMG network. Yahoo has
an exponential decay for intermediate degrees of the $P_{c}^{in}(k)$ distribution
[Fig.\ref{fig:fig02}(c)], although
it is lost for both low/high degrees. 
When looking at $P_{c}^{out}(k)$
[Fig.\ref{fig:fig02}(d)], we see
that the highest number of nodes is limited to 40. Furthermore, it exists a typical
number of $\sim30$ outgoing links of each network, as we infer from the constant
value (close to one) of $P_{c}^{out}(k)$ from $k=1$ to $k\sim30$. A similar characteristic
was shown at the $P_{c}^{out}(k)$ of MSN network [Fig.\ref{fig:fig02}(d)] (in that case, outgoing
links where set to $k_{out}=5$).

\section{DISCUSSION}

Some conclusions are drawn from the analysis of the networks' structure. Roughly
speaking, we can say that we observe two types of network, one with
power law distribution of $P_{c}^{in}(k)$  and the second
with an exponential decay.
Since all networks are supposed to have the same functionality, i.e. recommendation of music based on artist similarities, it would be expected that all of them have the same kind
of structure, nevertheless MSN and Amazon are scale-free while Yahoo and AMG are exponential.

The reason of the structural differences could be explained by taking into
account the ground characteristics of music networks together with their linking criteria.
From the point of view of network categories ~\cite{new03}, music networks 
can be considered as an information (or knowledge) network \cite{maslov} with a high social component. 
The structure of citations between academic papers is a classic example of 
an information network~\cite{ren98}.
In this sense, when an artist plays similar music 
to other artists, is somehow ``citing'' their music. At the same time, a social component
is unavoidable, since people are the elemental nodes of music networks. 
In both cases, social and information networks, scale-free structure has been 
reported ~\cite{new03}.
This structure is associated  with the preference of new nodes to associate
with nodes with high degree, i.e. with a high number of links ~\cite{bar99}. 

MSN and Amazon are user-ratings and user-habits based networks. In both
cases, links between artists are created by
user ratings, buying behaviour or downloads statistics obtained from thousands of users in what is 
known as ``collaborative
filtering''~\cite{sar01}. In such setups, each user inputs information on some of the nodes.
Then all the information is aggregated and combined with the aim of
predicting future ratings or, as explained in~\cite{sar01},
to calculate the similarity between nodes.
Of course, users have higher probabilities of linking artists (by rating or downloading) that he/she likes. Since some artists are much more popular than others, they will get more links.
From the obtained results, we find that this kind of collaborative filtering leads to scale-free structure, at least for the case of music recommendation networks. 

\begin{figure}[!h]
\centering
\includegraphics[angle=0, width=120mm,clip]{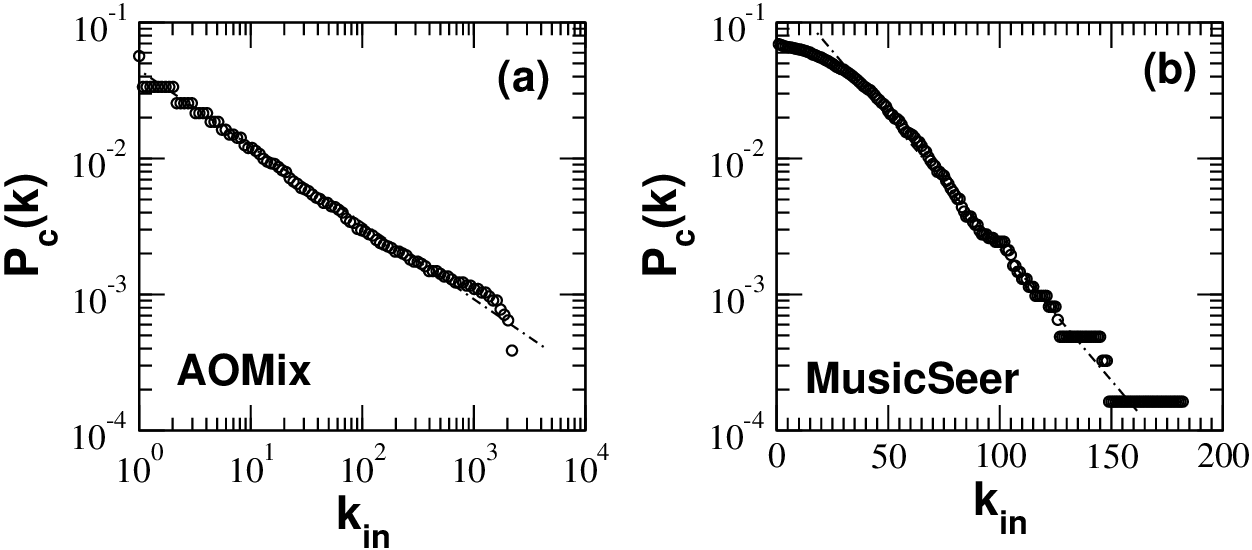}
\caption{
$P_c(k)$ of ArtOfTheMix (a) and MusicSeer (b) networks, in log-log and linear-log scale respectively.
Only the undirected distributions are displayed for ease of
comparison (the ArtOfTheMix graph is undirected).
}
\label{fig:fig03}
\end{figure}

 In order to check the hypothesis 
that popularity is behind the scale-free distribution, we use 
 user behaviour information from Art of the Mix~\cite{playlist}. Art of the Mix (AOMix) is
 a website dedicated to sharing of playlists submitted by a community
 of users. It contains almost 100,000 playlists contributed by
 thousands of users. Playlist information from AOMix has been previously used
 by~\cite{Berenz00, ell02} as a source for music artist similarity. The
 underlying assumption considers that artists that co-occur in the same playlist
 are somehow similar. This is the idea behind ``people
 who bought X also bought Y'' commonly found on on-line retailers such as Amazon. 
 The properties of the network constructed adding a link between any two artists
 that coincide in a playlist are depicted in Table~\ref{tab:tab02}. The cumulative degree distribution of AOMix is displayed on Fig.~\ref{fig:fig03} and shows a power-law decay
as MSN and Amazon. 
Art of the Mix is originally a bipartite graph composed of playlist and artists nodes that has been projected 
into an artist nodes graph.  It is worth mentioning that all networks built with collaborative-filtering algorithms 
derive from originally bipartite networks: e.g.: People who listened to this {\em X} also listened {\em Y} or People who liked {\em X} also liked {\em Y}. 
The use of what people do with or say about items, hence the exploitation of information of bipartite graphs, is the basis of collaborative filtering~\cite{lin03, sar01, maslov}.

Concerning the exponential scaling of Yahoo and AMG,
we have no information about the wiring mechanisms of Yahoo (due to their privacy policy).
Nevertheless, the AMG linking criterion is explained in detail in~\cite{all}. This recommendation
network is characterized  by an editorial group of 
``experts'' which supervises the wiring of the network. In this case, the construction of the network is uniquely guided
by similarity criteria, a fact that can not be guaranteed in the case of user-rating or user-behaviour derived networks. In addition, it could be expected that human experts are in
fact truncating possible scale-free by filtering links between normal
artists and ``hub'' artists~\cite{mos02}. 

Related with the exponential decay of Yahoo and AMG, it is worth commenting another network used by~\cite{ell02, Berenz00} in a pursue of a ground truth for music similarity. During a web experiment, named MusicSeer, users where asked to select
the most similar artist to a given one from a list of 10 possibilities. 
The properties of this human supervised network, where users have been
 explicitly asked to focus on similarity, is shown in Tab~\ref{tab:tab02} and
its degree distribution is depicted in Fig.~\ref{fig:fig03} (b). In this case, as for Yahoo and
AMG, an exponential degree distribution is obtained, and it is another
example of how similarity music networks which try to avoid user-preferences are
prone to have exponential decay. 

Both
networks' degree distributions, AOMix and MusicSeer, are
drawn next to each other to highlight how differences in the construction mechanisms 
affect the degree distribution. 
Exponential degree appears when similarity dominates over artist
popularity otherwise we observe power-law decays.

\begin{table*}[htb]
\begin{center}
\begin{tabular}{|r|c|c|c|c|c|c|}\hline
& type       &  $n  $  &$m  $    &  $\left<k\right>$ &  $C$ &   $C_r$   \\
\hline \hline
ArtOfTheMix
&undirected& 48,170 & 300,708 & 12.5 & 0.1 & 0.003  \\
\hline
MusicSeer
&directed& 6,144 & 10,219 & 2.9 & 0.02 & $4.7 \centerdot 10^{-4}$   \\
\hline
\end{tabular}
\small
\caption{Summary of the network parameters for the Art of the Mix and
the MusicSeer networks.}
\label{tab:tab02}
\end{center}
\end{table*}
\normalsize

A crucial issue of the quality of recommendation networks is how searchable they networks are, 
that is, how easy is it for a user to find a target quickly. The influence of the network structure on the navigation has been addressed
in the literature~\cite{kle00,wat02,mou03,kim02,ada01,men02}. Strategies for search in scale-free networks using local information
have been proposed by~\cite{kim02,ada01}. The algorithm selects the nodes with highest degree and scales sublinearly 
with the number of nodes. This type of algorithm cannot be exploited in the networks at hand. 
Firstly, the power law is found for the in-degree distribution only. As we mentioned above, 
the out-degree distribution has a cut due to web page usability constraints; the recommendation should fit on a web page. 
The in-degree distribution is unknown to the users, so it is unlikely that they choose the above proposed algorithm 
when searching in the recommendation networks. Secondly, it is unrealistic to think that users of the network 
can adopt such a search strategy but rather the selection of nodes will be guided by their intrinsic qualities or some sort of underlying distance~\cite{wat02,men02,mou03}. 
Kleinsberg showed that lattices with random  long-range links connected according to a distance dependent probability 
distribution are searchable~\cite{kle00}. Sublinear searches can be obtained assuming small-word regime and the 
existence of an distance between nodes~\cite{mou03}. Given its importance on the application,
 more work in this direction needs to be addressed. It will be very interesting to gather and analyzed statistics of navigation of real users of the system.

\section{CONCLUSIONS}

We have analyzed the structure of music recommendation networks by means of complex 
networks analysis. We have found small-world properties in all networks, which 
have an influence on the navigation properties of the network. Despite sharing
the small-world structure, we have found differences in the scaling of their
degree distribution. Networks
with user-preferences (from users of the network) as the linking criterion 
have a power-law decay of their degree distribution, i.e. show scale-free properties. 
On the other hand, networks constructed by similarity criteria, 
lead to an exponential decay of the degree distribution.
We believe that the scale-free and exponential decay
could be related with the social or information nature of the network. When
the music recommendation network is constructed under supervision of 
an editorial group, similarity aspects take advantage over the social ones and this
reflected in an exponential decay of the degree distribution. On the 
contrary, it is reasonable to expect that the social nature of the recommendation
network increases in case of user-preferences linking. Finally, we give some
insights about navigation through these kind of networks and address future work towards
this point.

\begin{acknowledgments}
We thank Juan A. Almendral, Fabien Gouyon and Pablo de Miguel for fruitful discussions.
Financial support was provided by 
MCyT-FEDER (Spain, projects BFM2002-04369 and BFM2003-07850), by the
Generalitat de Catalunya and by SIMAC IST-FP6-507142 European project.
\end{acknowledgments}


\begin{thebibliography}{99}
\bibitem{kmax} The value of $k_{max}$, which is the highest degree in the network, can
be estimated in scale-free networks as $k_{max}\sim n^{\frac{1}{\gamma-1}}$.
\end{thebibliography}
\end{document}